\documentclass[floatfix,showkeys,twocolumn,citeautoscript,showpacs,amsmath,amssymb,epsf,aps,prb]{revtex4}
\usepackage{dcolumn}
\usepackage{epsfig}

\usepackage{graphicx}
\usepackage{longtable}
\usepackage{dcolumn}
\usepackage{bm}

\def\trip_b{$^3$B$_1$ }
\def\sing_a{$^1$A$_1$ }
\def\af{$\alpha$ }
\def\afHF{$\alpha_{HF}$ }
\def\Xa{X$\alpha$ }
\def\rhup{\rho_{\uparrow}}
\def\rhdn{\rho_{\downarrow}}
\def\vrp{\vec{r'}}
\def\vr{\vec{r}}
\def\bo{\overline{\rho}^{\frac{1}{3}}}
\def\ro{\overline{\rho}}
\def\bt{\overline{\rho}^{\frac{2}{3}}}
\def\gbo{\overline{g}^{\frac{1}{3}}}

\def\gbt{\overline{g}^{\frac{2}{3}}}
\def\ss{\sigma}

\begin{document}

 \title{ Slater's exchange parameters  \af for Analytic and Variational \Xa calculations}

\author{Rajendra R. Zope}
\email{rzope@alchemy.nrl.navy.mil}
\affiliation{Department of Chemistry, George Washington University, Washington DC, 20052, USA}
\altaffiliation{Mailing address: Theoretical Chemistry Section,   Naval Research
Laboratory, Washington DC 20375-5345, USA}

\author{Brett I. Dunlap}
\email{dunlap@nrl.navy.mil}
\affiliation{Code 6189, Theoretical Chemistry Section, US Naval Research Laboratory
Washington, DC 20375}

\date{\today}

\begin{abstract}
  Recently, we formulated a fully analytical and variational implementation 
 of a subset of density functional theory using Gaussian basis sets to 
 express orbital and the one-body effective potential. The implementation, called 
 the Slater-Roothaan (SR) method,  is an extension 
 of Slater's \Xa method, which allows arbitrary scaling of the exchange potential 
 around each type of atom in a heteroatomic system. The scaling parameter
 is Slater's exchange parameter, $\alpha,$ which can be determined for each type of atom 
 by choosing various criteria depending 
 on the nature of problem undertaken. Here, we  determine these scaling 
 parameters for atoms H through Cl by constraining some physical quantity
 obtained from the self-consistent solution of the SR method to be equal
 to its {\em exact} value. Thus, the sets of \af values that reproduce the 
 {\em exact} atomic energies have been determined for four different combinations
 of basis sets.  A similar set of \af values that is independent of basis set is 
 obtained from numerical calculation. These sets of $\alpha$  parameters are subsequently
 used in the SR method to compute atomization energies of the G2 set of molecules. 
 The mean absolute error in atomization energies is about 17 kcal/mol, and is 
 smaller than that of Hartree-Fock theory (78 kcal/mol) and the local density 
 approximation (40 kcal/mol) but, larger than that of a typical generalized 
 gradient approximation ($\sim$ 8 kcal/mol).  A second set
 of \af values is determined by matching the highest occupied eigenvalue of the SR
 method to the negative of the first ionization potential.  Finally, the  possibility of
 obtaining \af values from the exact atomization energy of homonuclear diatomic molecules 
 is explored.  We find that the molecular \af values
 show much larger deviation than what is observed for the atomic \af values.
 The \af values obtained for atoms in combination with analytic SR 
 method allow elemental properties to be extrapolated to heterogeneous molecules.
 In general, the sets of different \af values might be useful for calculations of different 
 properties using the analytic and variational SR method.
\end{abstract}

\pacs{ }


\maketitle

    The Hohenberg-Kohn-Sham (HKS) formulation of density functional theory is by far
the choice for today's electronic structure calculation \cite{HK,KS65,Slater51,ratner}.
Prior to the HKS formulation of the density functional theory (DFT),
Slater formulated its basis in an attempt to reduce the computational complexity 
of the Hartree-Fock (HF) method\cite{Slater51}.  Unlike the HF method, this 
\Xa method
has an exchange potential $v_x$  that is local, and 
proportional to the one-third power of the electron density $\rho$. It is given 
by 
$$ v_x [\rho] =  - \alpha \frac{3}{2}
 \Bigl( \frac{3}{\pi} \Bigr )^{1/3} \rho^{1/3}(\vec{r}) ,$$ 
where $\alpha$ is called Slater's exchange parameter and 
was originally equal to one.
G\'asp\'ar \cite{Gaspar} and Kohn-Sham \cite{KS65}  obtained similar 
expression on other rigorous grounds. They variationally minimized 
the total energy functional and determined \af  to be  $2/3$,
in contrast to Slater's value.  The difference in the two values 
is rooted in the averaging process employed in the simplification of the HF  
exchange potential. Slater, taking a cue from Dirac's earlier work\cite{Dirac30}, averaged the 
exchange potential over the entire Fermi sphere of radius 
$k_f = (3\pi^2\rho(\vec{r}))^{1/3}$ while in the G\'asp\'ar's and Kohn-Sham 
procedure the value of exchange potential  at $k = k_f(\vec{r}) $ is used.
It was suggested that the \af parameter in Slater's method could be treated 
as  an adjustable parameter.\cite{Xalpha1}  The set of \af parameters for atoms that make
self-consistent \Xa energy  match the HF  energy was determined by 
Schwarz\cite{Schwarz72}.
The \af values he obtained typically  range from 0.77 for light atoms to 
0.69 for heavy atoms.  He also noted that different atomic configurations
lead to only slight changes in the \af values,  suggesting  that the \Xa
method can also be applied to the molecules or solids.  Several other 
ways to determine the \af parameters have been put forth.\cite{Smith70,Smith78,a1,a2,a3,a4,a5,OOS2000}

 Early implementations of the \Xa method for molecules with atom-dependent \af parameters
employed the muffin tin (MT) approximation\cite{MT1,MT2,MT3,MT4,MT5}.
In this scheme, atoms or ions are enclosed by 
the atomic spheres in which the potential is approximated to be spherically symmetric
while, in the interstitial region, the potential is constant. 
The effective one body potential in the MT implementation is discontinuous 
at the surface of MT sphere and  therefore the energy was mathematically undefined.
This method, however, had an advantage over all the other quantum-chemical
methods which was that the molecules dissociated correctly.  At 
infinite inter-atomic distances a sum of  atomic energies could be reproduced. When 
the MT model
was discarded due to difficulties in geometry optimization, a single
value of \af ($ \, = 0.7$) had to be chosen for all types of atoms.\cite{Barends}
There are, however, some attempts to obtain good thermo-chemistry 
from the X$\alpha$ calculations by  correcting or improving the uniform $\alpha$ calculations 
in a secondary calculation.\cite{Fliszar1,Fliszar2,Fliszar3,Fliszar4}

   Electronic structure calculations using traditional quantum chemical methods 
such as the HF theory, and beyond, generally employ basis sets to expand 
molecular orbitals.  In these methods, calculations of matrix 
elements and other quantities of interests are analytic. On the other hand,
almost all implementations of density functional models, until recently, required use 
of numerical grids to compute contributions to matrix elements from the exchange-correlation 
terms. Cook and coworkers\cite{Dunlap79,Cook86,Cook95,Cook97,BID89}
successfully demonstrated fully analytic implementation of density functional model.
His variational implementation used Gaussian basis sets to express molecular orbitals,
the one-body effective potential\cite{Dunlap79}, and the  uniform \af \Xa exchange
potential\cite{Cook95, Cook97}.
The advantages of analytic calculations are obvious. The calculations are fast and 
accurate to machine precision (within the limitations  of the model and basis set). 
Round-off error, which grows as the square root of number of points 
and other numerical problems are eliminated\cite{Boys}.
Consequently, smooth potential energy 
surfaces are obtained\cite{Cook95,Cook97}. 
Unfortunately, modern sophisticated exchange-correlation functionals  are too complex to 
allow fully analytic solution at this time. Analytic DFT, at this stage,
is restricted to Slater's exchange-type  functionals.  We have recently proposed
an algorithm  that permits fully analytic solutions and also allows for atom dependent 
values of Slater's exchange parameter \af for heteroatomic systems\cite{Dunlap03}. 
This method, called the Slater-Roothaan
(SR) method, is based on robust and variational fitting and requires four sets of Gaussian bases.
It can have the advantage of MT \Xa that molecules dissociate correctly as interatomic distances tend 
to infinity.\cite{Jhonson75} Intuitively, it is natural to expect that when molecules dissociate the constituent 
atoms  would have {\em exact} atomic energies at infinite separation.  This important physical 
requirement is satisfied within the SR method if  \af parameters that reproduce exact 
atomic energies are used in molecular calculations, although one might want to use
other sets of $\alpha's$ for other molecular properties.

   In this work, we report different sets of \af parameters that could be used in the
SR method. The first such a set is determined by requiring that the self-consistent 
atomic  SR atomic energy be equal to the {\em exact} atomic energies. The \af values in 
this set will be hereafter referred to as atomic alpha values. We have recently 
used some of the \af values from this set to calculate the {\it total} energies 
for the G2 set\cite{Pople89} of molecules and found that these \af values give remarkably 
good total energies.\cite{ZB_PRB05} Here, we use them to compute the {\it atomization}
energies  for the G2 
set of molecules and compare them with other models. 
Further, we also obtain two additional sets of \af based on other criteria for possible
use in the SR method.
The second set of atomic \af  values is determined to provide negative of eigenvalue of the
highest occupied molecular orbital (HOMO), $\epsilon_{HOMO}$, that matches with the {\em exact}
value of the first ionization potential are 
determined.\cite{ehomo1,ehomo2,ehomo3,ehomo4,ehomo5,ehomo6}
  The third set of \af  is determined to provide  the {\em exact}
atomization energy for the selected homonuclear diatomic molecules of the first and second
row of the periodic table. All calculations are repeated for four different combinations 
of analytic Gaussian basis sets.

    It is apparent that the analytic Slater-Roothaan method is empirical in nature as it 
uses adjustable \af values. It,  however,
differs from the semi-empirical\cite{S1,S2,S3} methods which use a minimal
basis set, or empirical\cite{E1,E2,E3,S4}  models which do not compute electronic structure,
and the minimal-basis tight binding methods\cite{T1} reported in the literature.
It is a variant of density functional models that is also computationally very efficient because 
it is fully analytic and requires {\em no} numerical integration.
We have successfully used it for studying the heterogeneous systems
such as boron and aluminum nitride nanotubes containing up to  200 atoms\cite{BN,AlN}.  
Furthermore,  it also provides the flexibility of being tuned, through the \af parameters,
according to the need of problem.
 The sets of \af parameters reported in this work will be  useful starting points
in this regard.
In the following 
section we outline the analytic SR method used in this work. In section II, we describe 
computational details which are followed by results and discussion in section III.

\section{Theoretical Method}

                The total electronic energy in the DFT for an $N$-electron 
system is a functional of electronic density $\rho$ and is   given by
\begin{equation}
E^{KS} [\rho]  = \sum_i^N <\phi_i| f_1| \phi_i> + E_{ee} + 
            E_{xc}\big [ \rhup, \rhdn \big] \label{eq:I}
\end{equation}
 where,  the first term contains the kinetic energy operator and the nuclear attractive potential due to
the $M$ nuclei,
\begin{equation}
f_1 = -\frac{\nabla^2}{2} - \sum_A^M \frac{Z_A}{\vert\vr - \vec{R_A}\vert}.
\end{equation}
The total electron density is expressed in terms of the Kohn-Sham orbitals $\phi_{i,\sigma}$ as
\begin{equation}
      \rho(\vr) = \rho_{\uparrow} +  \rho_{\downarrow},
\end{equation}
with
\begin{equation}
      \rho_{\sigma}(\vr) = \sum_{i} n_{i,\sigma} \phi_{i,\sigma}^* \phi_{i,\sigma}(\vr),
\end{equation}
where $n_{i,\sigma}$ is the occupation number for the $\phi_{i,\sigma}$ orbital.
The  second term in Eq. (\ref{eq:I}) represents the classical Coulombic interaction energy
of electrons,
\begin{equation}
  E_{ee} = \langle  \rho\vert\vert\rho \rangle  =
    \frac{1}{2} 
 \int  \int \frac{\rho(\vr)\rho(\vrp)}{\vert\vr-\vrp\vert} d^3r \, d^3r'.
\end{equation}
This energy is  approximated  by expressing the  charge density as a fit to a set of Gaussian functions,
\begin{equation}
\rho(\vr) \approx \overline{\rho}(\vr) = \sum_i d_i G_i(\vr),
\end{equation}
where, $\ro (\vr)$ is the fitted density, d$_i$ is the expansion coefficient of the charge 
density Gaussian basis function G$_i$. The elimination of the first order error in the total energy due to the fit
leads to\cite{Dunlap79}
\begin{equation}
 E_{ee} = 2 \langle {\ro}\vert\vert\rho \rangle  -  
            \langle {\ro}\vert\vert{\ro} \rangle .
\end{equation}
The expansion
coefficients $\{d_i\}$ are determined by variationally minimizing this energy with respect to $\{d_i\}$.
 The last term E$_{xc}$ in Eq. (\ref{eq:I}) is the exchange energy, 
\begin{equation}
E_{xc} [\rhup, \rhdn]  =  - \frac{9}{8} \alpha \Big ( \frac{6}{\pi} \Big )^{1/3} \int d^3r 
            \Big [ \rhup^{\frac{4}{3}} (\vr) +  \rhdn^{\frac{4}{3}} (\vr) \Big]. \label{eq:Exc}
\end{equation}
  The form of above functional allows analytic calculations with the Gaussian basis to be performed.
For this purpose the one-third and two-third powers of the electron density are also expanded in 
Gaussian basis sets: 
\begin{eqnarray}
            \rho^{\frac{1}{3}} (\vr) \approx \bo =  \sum_i e_i  E_i (\vr)  \\
            \rho^{\frac{2}{3}} (\vr) \approx \bt =   \sum_i f_i  F_i (\vr) .
\end{eqnarray}
 Here, $\{E_i\}$ and $\{F_i\}$ are independent Gaussian basis functions, while $e_i$ and $ f_i$ are expansion coefficients.
The exchange energy is then given by\cite{Dunlap86,Cook86,BID89,Cook95,Cook97}
\begin{equation}
E_{xc}   =  C_{\alpha} \Bigl [ \frac{4}{3} \langle \rho \,  \bo \rangle - \frac{2}{3} \langle \bo  \, 
     \bo  \, \bt \rangle
    + \frac{1}{3} \langle \bt \, \bt \rangle 
 \Bigr ] ,
\end{equation}
 where $C_{\alpha} = - {9} \alpha \Big ( \frac{3}{\pi} \Big )^{1/3} $
  Thus, using the four LCGO basis sets (one for orbital expansion and three fitting basis sets) the 
total energy is calculated analytically.
The LCAO orbital coefficients and the vectors {\bf d}, {\bf e}, and {\bf f} are found 
by (constrained if desired) variation.

\subsection{Slater-Roothaan method}
   The expression for the total electronic energy in the Slater-Roothaan method has the following
 form:\cite{Dunlap03}
\begin{eqnarray}
  E^{SR} &  = & \sum_i <\phi_i| f_1| \phi_i> + 2 \langle \rho \vert\vert\ro\rangle 
                - \langle \ro\vert\vert\ro\rangle    \nonumber \\
          & & \,\, + \sum_{\sigma = \uparrow, \downarrow}  
             \Bigl   [
            \frac{4}{3} \langle g_{\ss}  \, \gbo_{\!\ss} \rangle
           - \frac{2}{3} \langle {\gbo}_{\!\ss} \,  {\gbo}_{\!\ss}  \, {\gbt}_{\!\ss} \rangle  \nonumber \\
         & & \,\, \, +  \,\frac{1}{3} \langle {\gbt}_{\!\ss} \,  \, {\gbt}_{\!\ss} \rangle \Bigr ].
\end{eqnarray}
  Here,  the partitioned $3/4$ power of the exchange energy density,
\begin{equation}
  g_{\ss} (\vr) =  C_x \sum_{ij} \alpha(i) \, \alpha(j) \, D_{ij}^{\ss} (\vr),
\end{equation}
where  $C_x = C_{\alpha}/{\alpha}$ and  $D_{ij}^{\ss} (\vr)$ is the diagonal part of the 
spin density matrix, and the function,
\begin{equation}
   \alpha(i) =  \alpha_i^{3/8}
\end{equation}
contains the $\alpha_i$, the \af in the \Xa, for the atom on which the atomic orbital $i$ is the centered.

\begin{table}
\begin{ruledtabular}
\caption{The optimal $\alpha$ values for the different basis sets that yield the 
{\em exact} atomic energies in the highest symmetry for which the solutions
have integral occupation numbers. The numerical values are for the 
spherically symmetric atoms and are obtained by the numerical  atomic structure code.
The {\em exact} atomic energies given in the last column are from Ref. \onlinecite{clemmenti}.
The basis sets are I: 6-311G**/RI-J, II: 6-311G**/A2, III: DGDZVP2/RI-J, and 
 IV: DGDZVP2/A2.}
\label{table:aalpha}
\begin{tabular}{llllllr}
  &  Basis I     & Basis II   & Basis III & Basis IV &  Numerical &  E (a.u.)      \\
\hline
  H &   0.77739  &  0.77739  &  0.78124  &   0.78124 &   0.77679  &     -0.500     \\
 Li &   0.79169  &  0.79169  &  0.79211  &   0.79211 &   0.79118  &     -7.478     \\
 Be &   0.79574  &  0.79574  &  0.79614  &   0.79614 &   0.79526  &    -14.667     \\
  B &   0.78675  &  0.78668  &  0.78684  &   0.78677 &   0.78744  &    -24.654     \\
  C &   0.77677  &  0.77672  &  0.77670  &   0.77665 &   0.77657  &    -37.845     \\
  N &   0.76747  &  0.76747  &  0.76726  &   0.76726 &   0.76654  &    -54.590     \\
  O &   0.76500  &  0.76495  &  0.76454  &   0.76448 &   0.76454  &    -75.067     \\
  F &   0.76066  &  0.76067  &  0.76002  &   0.76001 &   0.75954  &    -99.731     \\
 Na &   0.75204  &  0.75204  &  0.75287  &   0.75287 &   0.75110  &   -162.260     \\
 Mg &   0.74994  &  0.74994  &  0.75120  &   0.75120 &   0.74942  &   -200.060     \\
 Al &   0.74822  &  0.74819  &  0.74872  &   0.74869 &   0.74797  &   -242.370     \\
 Si &   0.74539  &  0.74540  &  0.74600  &   0.74602 &   0.74521  &   -289.370     \\
  P &   0.74324  &  0.74324  &  0.74397  &   0.74397 &   0.74309  &   -341.270     \\
  S &   0.74262  &  0.74260  &  0.74352  &   0.74350 &   0.74270  &   -398.140     \\
 Cl &   0.74197  &  0.74196  &  0.74273  &   0.74272 &   0.74183  &   -460.200     \\
\end{tabular}
\end{ruledtabular}
\end{table}

\section{Computational Details}

      The \af values are calculated for the atoms that occur in standard set of (Becke's \cite{Becke93})
56 molecules. These atoms are H, Be through F, and Na through Cl. Magnesium and aluminum 
atoms do not occur in the Becke's set but have been included in the present work. The set of 56
molecules  actually consist of 54 molecules with two molecules in two different electronic states.
The SR method requires four basis sets, of which one is the orbital basis.
The remaining three basis sets are required for fitting 
of the the Kohn-Sham potential (See  Ref. ~\onlinecite{Dunlap03} for
more details).  For the orbital basis sets, our choices are 
the triple-$\zeta$ (TZ) 6-311G** basis\cite{O1,O2} and the DGauss\cite{AW91} valence
double-$\zeta$  (DZ) basis set\cite{GSAW92} called DZVP2.  
The most reliable and thus best {\sl s-}type fitting bases are those that are 
scaled from the {\sl s-} part of the orbital basis \cite{Dunlap79}. 
The scaling factors are two for the density, $\frac{2}{3}$ for $\bo$ \,\, and $\frac{4}{3}$ 
for $\bt$.
These  scaled bases are used for all {\sl s-}type fitting
 bases.  A complete package of basis sets has been optimized \cite{GSAW92} 
for use with DGauss \cite{AW91}.  
In addition to the valence double-$\zeta$ orbital basis, called DGDZVP2 herein, we use
the {\sl pd} part of the (4,3;4,3) for Be-F (A2) charge density fitting basis.
Ahlrichs' group has generated a RI-J basis for fitting the charge density of a valence 
triple-$\zeta$ orbital basis set used in the {\sc Turbomole}
program \cite{EWTR97}.  These fitting bases are used in combination with the 6-311G** 
and DGDZVP2 basis sets. The \af are obtained for the different combination of  above basis
sets. The molecule geometries were optimized using the Broyden-Fletcher-Goldfarb-Shanno (BFGS) 
algorithm\cite{BFGS1,BFGS2,BFGS3,BFGS4,BFGS5}.
Forces on ions are rapidly computed non-recursively using the 4-j generalized Gaunt 
coefficients \cite{Dunlap02, Dunlap05}. The calculations are spin-polarized for open-shell 
systems.

\section{Results and Discussion}

The optimal \af  values that give the exact atomic energies can be  obtained by 
the Newton-Raphson procedure that finds zeros of the function
 $ f(\alpha) =  E^{SR} (\alpha) - E_{exact} = 0, $ 
where $E$ and $ E_{exact}$ are the self-consistent \Xa (SR)  and the {\em exact} total 
energies, respectively. 
Finding zero of this function requires 
frequent calculation of this  function and its derivative. Determination of derivative,
$f^{\prime}(\alpha),$ is 
straight forward. Only the exchange term in the energy functional $E^{SR}(\alpha)$ 
depends explicitly on the \af parameter. 
The derivative obtained as the exchange-energy divided by the $\alpha$ parameter 
provides sufficiently accurate approximation to the actual derivative. 
The LCAO fits depend weakly 
on $\alpha$.  The Newton-Raphson procedure 
was implemented using PERL scripts and the energy and its derivative with respect 
to \af were obtained from the FORTRAN90 SR code. The sets of \af values obtained using this procedure 
are given in Table in \ref{table:aalpha}.
It is evident from the table that the choice of fitting  basis does not significantly affect
the  \af values. The \af values do show some dependence on the orbital basis set. 
However, the changes in alpha values are small  with the \af values for the 6-311G** basis set being
consistently smaller than those for the DGDZVP2 basis set for the same fitting basis set.
For the same choice of fitting basis sets, the 6-311G** orbital basis set will give lower
energy than the  DGDZVP2 basis. Therefore, the \af values for the DGDZVP2 basis should be larger
than the 6-311G** basis  to provide the more negative (or binding) potential required 
to give the energy equal to the {\em exact} energy. 
Except for H, which has no correlation energy, 
these \af values are larger than the reported HF  values\cite{Schwarz72} which 
is expected as these values  are obtained for the exact energy (bounded from above by the HF energy).
The HF \af values show systematic  monotonic decrease with 
the atomic number\cite{MT3}.
  The current \af values  show overall decrease with the atomic
number except for peaks at Li and Be.
The peak at Li is caused by the fact that H has no correlation energy
while peak for Be is caused by electron pairing and correlation in the outer orbital.

\begin{table}
\begin{ruledtabular}
\caption{The optimal $\alpha$ values for the different basis sets for which 
the eigenvalue of the highest occupied molecular orbital is   exactly equal to the {\em exact} ionization 
potential (IP). The experimental values of the first ionization  potentials (in eV) are also included in the 
last column.  The basis sets are I: 6-311G**/RI-J, II: 6-311G**/A2, III: DGDZVP2/RI-J, and 
 IV: DGDZVP2/A2.}
\label{table:aehomo}
\begin{tabular}{lllllr}
  &  Basis I     & Basis II    & Basis III &   Basis IV  &   IP\cite{Moore}      \\
\hline
  H &   1.1901  &  1.1901  &  1.1877  &   1.1877  &   13.60 \\ 
 Li &   1.1246  &  1.1246  &  1.1152  &   1.1152  &    5.39 \\ 
 Be &   1.2749  &  1.2749  &  1.2581  &   1.2581  &    9.32 \\ 
  B &   1.1227  &  1.1220  &  1.0993  &   1.0989  &    8.30 \\ 
  C &   1.1006  &  1.1000  &  1.0775  &   1.0770  &   11.26 \\ 
  N &   1.0832  &  1.0832  &  1.0629  &   1.0629  &   14.52 \\ 
  O &   1.2360  &  1.2345  &  1.1958  &   1.1946  &   13.61 \\ 
  F &   1.1696  &  1.1694  &  1.1501  &   1.1498  &   17.42 \\ 
 Na &   1.1181  &  1.1181  &  1.1221  &   1.1221  &    5.14 \\ 
 Mg &   1.2256  &  1.2256  &  1.2311  &   1.2311  &    7.64 \\ 
 Al &   1.1113  &  1.1112  &  1.1074  &   1.1067  &    5.98 \\ 
 Si &   1.0864  &  1.0836  &  1.0875  &   1.0844  &    8.15 \\ 
  P &   1.0501  &  1.0501  &  1.0662  &   1.0662  &   10.49 \\ 
  S &   1.1543  &  1.1562  &  1.1759  &   1.1770  &   10.35 \\ 
 Cl &   1.1211  &  1.1218  &  1.1344  &   1.1347  &   12.96 \\ 
\end{tabular}
\end{ruledtabular}
\end{table}

      The atomic \af values can also be determined by  different criteria.
In the KS DFT it has been argued that the negative of the eigenvalue of the highest 
occupied orbital $-\epsilon_{max}$ equals the first ionization 
potential.\cite{ehomo1,ehomo2,ehomo3,ehomo4,ehomo5,ehomo6} 
Although such an interpretation of the $-\epsilon_{max}$ is not 
yet settled, it gives us another way to determine set of  \af parameters.
Such a such set of $\alpha's$  might be useful in future for calculations 
of polarizability or optical spectra.
These \af values are given in  Table \ref{table:aehomo} for the four combinations of basis set.
All the values presented are larger than those obtained from the total energy 
matching criteria. This is not surprising as
the $-\epsilon_{max}$ for the local density type ($\alpha=2/3$) functionals 
typically  underestimate the ionization potential by as much as 30-50\%.  It is known that this occurs 
due to the incorrect asymptotic behavior of the effective potential. The asymptotic
behavior of the effective potential is governed by the exchange potential. In the 
present \Xa method, the exchange potential decays  exponentially in the asymptotic
region, due to which the valence electrons experience shallower long-range potential than they 
otherwise should.  In the present work, we, however, treat \af as purely a fitting parameter
to get the right $\epsilon_{max}$.   
These \af values do not decrease left to right across the periodic table and 
are rather close to Slater's value of $\alpha$. The removal of self-interaction of electron
also leads to better asymptotic description\cite{PZ81}. We think that the 
self-interaction correction can be also be implemented in the present analytic SR method.
Orbital densities are non-negative.

\begin{table}
\begin{ruledtabular}
\caption{The optimal $\alpha$ values for the selected homonuclear dimers that yield the
{\em exact} atomization energies for the 6-311G**/RI-J basis set. The {\em exact} formation
 energies and the bond lengths (in \AA)  are also included in the last column.}
\label{table:malpha}
\begin{tabular}{lcrcc}
        &  $\alpha$   &  $D_0$ (kcal/mol)  & $R_e$ & $R_e$       \\
        &             &           & present &  Expt.\cite{Huber79}    \\
\hline
 H$_2$  &   1.39172  &    -103.5  & 0.59  & 0.74  \\
 Li$_2$ &   1.45747  &     -24.0  & 2.12  & 2.67     \\
 Be$_2$ &   0.56596  &      -2.4  & 2.82  & 2.46     \\
 C$_2$  &   0.81530  &    -148.7  & 1.21  & 1.24    \\
 N$_2$  &   0.88901  &    -225.1  & 1.04  & 1.10 \\
 O$_2$  &   0.42790  &    -118.0  & 1.36  & 1.21 \\
 F$_2$  &   0.29685  &    -36.9   & 1.67  & 1.41       \\
 Na$_2$ &   1.61278  &    -16.6   & 2.27  & 3.08       \\
 Si$_2$ &   0.78366  &    -74.0   & 2.10  & 2.24       \\
 P$_2$  &   1.43361  &    -116.1  & 1.58  & 1.89 \\
 S$_2$  &   0.64867  &    -100.7  & 1.96  & 1.89 \\
 Cl$_2$ &   0.66335  &     -57.2  & 2.06  & 1.99 \\
\end{tabular}
\end{ruledtabular}
\end{table}

\begin{table}
\begin{ruledtabular}
\caption{The atomization energies D$_0$ (kcal/mol) for the 56 set of molecules calculated using  the
two sets of $\alpha$ values given in Table I for the two basis sets 
that reproduce the {\em exact} atomic energies. 
The two  basis sets are I: 6311G**/RIJ, II: DZVP2/A2.
The last column contains the {\em exact} values. }
\label{table:m56}
\begin{tabular} {lrrr}
          &  Basis  I     &  Basis II  &         Exact \\
 H$_2$ 	 &  85.1 	 &  87.3 	 & 110.0\\ 
 LiH 	 &  38.1 	 &  33.8 	 &  57.7\\ 
 BeH 	 &  58.2 	 &  32.7 	 &  49.6\\ 
 CH 	 &  67.0 	 &  68.2 	 &  83.7\\ 
 CH$_2$($^3B_1$) 	 & 195.3 	 & 197.0 	 & 189.8\\ 
 CH$_2$($^1A_1$) 	 & 158.1 	 & 161.4 	 & 180.5\\ 
 CH$_3$ 	 & 299.2 	 & 302.4 	 & 306.4\\ 
 CH$_4$ 	 & 407.3 	 & 411.8 	 & 419.1\\ 
 NH 	 &  67.1 	 &  68.8 	 &  83.4\\ 
 NH$_2$ 	 & 157.1 	 & 161.5 	 & 181.5\\ 
 NH$_3$ 	 & 271.8 	 & 278.4 	 & 297.3\\ 
 OH 	 &  98.6 	 & 100.0 	 & 106.3\\ 
 H$_2$O 	 & 228.0 	 & 232.8 	 & 232.1\\ 
 HF 	 & 144.8 	 & 148.4 	 & 140.7\\ 
 Li$_2$ 	 &   6.6 	 &   5.6 	 &  24.4\\ 
 LiF 	 & 147.7 	 & 136.4 	 & 138.8\\ 
 C$_2$H$_2$ 	 & 422.8 	 & 414.2 	 & 405.3\\ 
 C$_2$H$_4$ 	 & 571.6 	 & 573.5 	 & 562.4\\ 
 C$_2$H$_6$ 	 & 711.7 	 & 718.1 	 & 710.7\\ 
 CN 	 & 190.5 	 & 181.4 	 & 179.0\\ 
 HCN 	 & 317.7 	 & 307.8 	 & 316.3\\ 
 CO 	 & 283.2 	 & 270.5 	 & 259.2\\ 
 HCO 	 & 307.4 	 & 301.4 	 & 278.3\\ 
 H$_2$CO (formaldehyde) 	 & 394.4 	 & 391.2 	 & 373.4\\ 
 H$_3$CO$_{}$H  	 & 521.9 	 & 527.0 	 & 511.6\\ 
 N$_2$ 	 & 215.0 	 & 203.6 	 & 228.5\\ 
 N$_2$H$_4$ 	 & 413.6 	 & 424.9 	 & 437.8\\ 
 NO 	 & 163.2 	 & 155.3 	 & 152.9\\ 
 O$_2$ 	 & 157.5 	 & 154.0 	 & 120.4\\ 
 H$_2$O$_2$ 	 & 283.3 	 & 289.0 	 & 268.6\\ 
 F$_2$ 	 &  66.5 	 &  68.3 	 &  38.5\\ 
 CO$_2$ 	 & 456.5 	 & 437.9 	 & 388.9\\ 
 SiH$_2$($^1A_1$) 	 & 125.4 	 & 128.0 	 & 151.4\\ 
 SiH$_2$($^3B_1$) 	 & 121.2 	 & 122.6 	 & 130.7\\ 
 SiH$_3$ 	 & 194.7 	 & 196.3 	 & 226.7\\ 
 SiH$_4$ 	 & 281.7 	 & 284.1 	 & 321.4\\ 
 PH$_2$ 	 & 126.1 	 & 130.4 	 & 152.8\\ 
 PH$_3$ 	 & 203.4 	 & 210.0 	 & 242.0\\ 
 H$_2$S 	 & 163.6 	 & 169.3 	 & 182.3\\ 
 HCl 	 & 101.4 	 & 102.5 	 & 106.2\\ 
 Na$_2$ 	 &   4.9 	 &   5.2 	 &  16.8\\ 
 Si$_2$ 	 &  72.0 	 &  72.2 	 &  74.7\\ 
 P$_2$ 	 &  94.1 	 &  94.5 	 & 117.2\\ 
 S$_2$ 	 & 110.2 	 & 111.7 	 & 101.6\\ 
 Cl$_2$ 	 &  64.4 	 &  65.9 	 &  57.9\\ 
 NaCl 	 &  88.8 	 &  90.6 	 &  97.8\\ 
 SiO 	 & 199.0 	 & 198.5 	 & 191.2\\ 
 CS 	 & 179.2 	 & 178.4 	 & 171.2\\ 
 SO 	 & 142.6 	 & 147.8 	 & 125.1\\ 
 ClO 	 &  79.5 	 &  85.6 	 &  64.3\\ 
 ClF 	 &  80.5 	 &  85.7 	 &  61.4\\ 
 Si$_2$H$_6$ 	 & 475.4 	 & 480.3 	 & 529.5\\ 
 CH$_3$Cl  	 & 400.5 	 & 404.8 	 & 393.6\\ 
 H$_3$CSH        	 & 467.8 	 & 475.8 	 & 472.7\\ 
 HOCl 	 & 174.5 	 & 182.1 	 & 164.3\\ 
 SO$_2$ 	 & 281.4 	 & 286.4 	 & 258.5\\ 
\end{tabular}
\end{ruledtabular}
\end{table}
\begin{table*}
\begin{ruledtabular}
\caption{ The mean and mean absolute error (kcal/mol) 
in the calculated {\it atomization} energy of 56 molecules (Cf. Table \ref{table:m56})
relative to their experimental values for different basis sets.
The SR (EA) are present calculations while SR (HF) are SR calculation with \afHF values.
The SR (HF) results are from Ref. ~\onlinecite{Dunlap03}.  The mean absolute error for the standard 
Hartree-Fock theory and the local density approximation is 78 and 40 kcal/mol\cite{Becke93,Curtiss97},
respectively.
OB: Orbital basis,  FB: Fitting basis (pdfg-type).  }
\label{table:err}
\begin{tabular}{lldddd}
   OB & FB &  SR (EA) & SR (EA) & SR (HF)  &   SR (HF) \\
   &              &       mean & absolute   & mean  & absolute \\
\hline
   6-311G**  &  RI-J   &  -1.9    &  17.3   & -5.1 & 16.4 \\
   DGDZVP2      &  A2   &   -1.6    &   16.2  & -4.8 & 16.1 \\
\end{tabular}
\end{ruledtabular}
\end{table*}

   Finally, we explore the possibility of obtaining exact atomization 
energy (AE)  within the present SR method, for selected diatomic molecules.
For this purpose we use Newton-Raphson procedure to obtain the  zero 
of the following function:
 $ f(\alpha) =  E^{SR}_{mol} (\alpha) - 2E^{SR}_{atom} (\alpha) - AE = 0, $ 
Here, $E^{SR}_{mol}$ is the self-consistent total energy 
of the optimized  molecules for a given $\alpha$, and $E^{SR}_{atom}$
is the self-consistent total atomic energy for a given alpha. The bond 
length of the molecule is optimized during each Newton-Raphson step.
These calculations were performed using PERL scripts and the calculated
set of \af values is presented in the Table \ref{table:malpha}.
These
values show large variation as function of atomic number. The general trend is
that the \af values for the dimers of atom on the left of periodic table
are larger than 1.0 and smaller than 1.0 for the atoms on the right. For the exact
values of the atomization energies the bond lengths of these dimers show
the general trend that the atoms on the left side of the periodic table
are larger and those on the right are smaller. The exception are the Be$_2$
and N$_2$ dimers, perhaps due to their weak and very strong bonding, respectively.
Another noteworthy observation is that these dimers with exception of Be$_2$
are still bound for vanishingly small value of \af.
This is interesting as no molecules are bound in the
Thomas-Fermi model.\cite{Teller62}  It seems the bonding therefore occurs due to
the exact treatment of kinetic energy in the present \Xa method. Here, it should be
noted that the 6-311G** basis sets used in this work may not provide satisfactory
description for  very small \af  values as the electrons then will experience a very
shallow potential and will be rather delocalized leading thereby to the artifacts
like positive eigenvalues for the outermost electrons.
For this reason the \af for the F$_2$ dimer should be used
cautiously.  
Further, these \af values when viewed relative to the Kohn-Sham's  value of
$2/3$ provide some hints on how the corrections to the Kohn-Sham exchange
should energetically contribute for  accurate determination of atomization 
energies. The molecules with molecular \af values larger than the Kohn-Sham value of $2/3$ 
will be  underbound in the exchange only KS scheme and those with \af 
lower will be overbound. Any universal correction to the KS exchange functional 
should be such that it simultaneously reduces overbinding in some molecules 
and underbinding in others.

   The atomization energies for the set of 54 molecules in 56 the electronic 
states are presented in the Table \ref{table:m56}. These are calculated by the
SR (EA) method using the set of \af values given in Table I. The calculations are 
performed for the two sets of basis sets. These molecules in the 
dissociation limit will give the {\em exact} atomized energies.
It is evident from the Table \ref{table:m56}
that the \Xa underbinds some molecules while it overbinds others. This trend
is in contrast with the local density approximation and the HF theory.
The former uniformly overbinds while the latter uniformly underbinds. A similar  trend has been 
found when the HF values of \af are used \cite{Dunlap03}. In general, the 
error in atomization energy is smaller when the molecules consist of 
atoms on the opposite sides of the periodic table, for example, HCl, H$_2$S,
NaCl, HO etc.  The best agreement is observed for the NH$_3$ with the DGDZVP2 
basis and OH with all basis.  The mean and mean absolute errors 
in the atomization energies of these molecules calculated with respect 
to their experimental values is given in Table \ref{table:err}.
These errors are  somewhat larger than those obtained when the HF \af values 
are used. We note here that the  HF  value of \af\!\!, \afHF\!\!,  are those that when used in the \Xa 
model reproduce the HF exchange energies for atoms.  

 The \af values obtained from the $\epsilon_{max}$ matching criteria 
can be used to obtain the first ionization energy of a molecule  from 
its highest occupied eigenvalue. Here, we demonstrate their use
for calculations of the first ionization potential of N$_2$ 
and CO molecules.  For the N$_2$ molecule,
using the \af values from Table ~\ref{table:aehomo} for the 
basis-I through basis-IV, the ionization potential is 
15.38,15.34, 15.40, and 15.20 eV, respectively. These values 
are in good  agreement with experimental value of 15.58 eV\cite{Curtiss_IP}.
For the CO molecule, the ionization potential is 14.83, 14.83, 14.53, and 14.53 eV 
for the basis sets I through IV. These values are also 
in very good agreement with CO's experimental 
ionization energy (14.01 eV)\cite{CO_IP}.
 We expect that use of these \af 
values to give better eigen-spectra, which could be used in the  
the  polarizability calculations by sum over states method\cite{SOS}. 
The atomic or Hartree-Fock $\alpha$ values could be used to optimize 
the geometry of molecule and then to calculate the matrix elements 
required in the sum over states expression. The eigenvalues required 
in the sum over states can be obtained by performing one more  
self-consistent calculations using the \af values from Table ~\ref{table:aehomo}.
The use of improved  eigen-spectrum in the sum over states method 
has been found to give good estimates of polarizability\cite{SOS}.
Calculations for the set of 54 molecules using molecular \af values 
(Cf. Table \ref{table:malpha}) are not  performed because the SR method as implemented
now\cite{Dunlap03} can not handle  wide variation in the molecular  
\af values.  
Moreover for the reasons mentioned earlier these \af  values
(particularly for the molecules containing $F$) necessitates 
bigger orbital basis sets.
The trend in the bond lengths in Table \ref{table:malpha} suggests
that for these dimers better values of {\af} may be obtained by 
minimizing the deviation of both the dissociation energy and the 
bond length, from their experimental values. There are several other 
possibilities for determination of \af which may be useful in 
improving  the performance of the SR method.
As mentioned earlier the differences in the Slater's and G\'asp\'ar-Kohn-Sham
value of \af are the result of different averaging process 
employed in the derivation of exchange potential.
This suggests the  possibility that different \af values for the 
description of the core (say $\alpha_c$) and valence ($\alpha_v$) electrons
may provide better description of the exchange correlations.  The additional parameter in this case 
will provide greater flexibility in the fitting procedure. Alternatively, 
the \Xa exchange functional can be augmented with suitable  
functional forms that allows the analytic solution of the problem. 
The local functional form by Liu and coworkers  that consists of the sum of 
$0, \, 1, \,1/3,$ and $2/3$ powers of electron density and originates from the 
adiabatic connection formulation appears to be particularly suitable 
for this purpose.\cite{LP96, LSLN96} We are currently exploring these possibilities.

  To summarize three sets of \af values are obtained on the different 
criteria for the use in the analytic SR method. The first set of \af
values is determined by equating the self-consistent total energy 
of atoms to the ``exact'' atomic energy. Two other sets are 
determined using different constraints such as equating the 
negative of eigenvalue of the highest occupied orbital to the first ionization 
potential and reproduction of exact atomization energies for the diatomic 
molecules. The examination of the performance of the SR method for 
the atomization energies of 56 molecules  with the first set of \af 
values gives  the mean absolute error to be about 17 kcal/mol. This 
value is far larger than the generally accepted chemical accuracy of 2 kcal/mol.
The tabulation of the MAE for the G2 set and extended G2 set for more 
sophisticated functionals is given in  Ref. \onlinecite{ES99}. The MAE  is within 
4-9 eV for the parameterizations at the level of the generalized-gradient 
approximation (GGA).\cite{PW91,PBE,BLYP,TVS98}  
The meta-GGA\cite{AES00} or hybrid GGA functionals\cite{Becke93,AB98} perform 
even better with MAE of about 3 eV. The MAE is 37-40 kcal/mol for the local spin 
density approximation\cite{Curtiss97}. The SR method's performance is 
intermediate between the local spin density approximation and the GGA.
 We note that we have not optimized the method to give accurate estimate of 
any particular property. The \af parameters could be chosen to minimize
the MAE in atomization energies like the other density functional models,
including the GGA, do. Such an optimization process would necessarily lower the 
error in Table ~\ref{table:err}.  Our goal in this work is to examine 
simple schemes for extrapolating elemental properties to heterogeneous 
molecules, without any additional massaging of the results.
For example, the set of \af values from Table \ref{table:aalpha} 
can be used to extrapolate atomic energies to molecular energies
by SR method. Indeed, such an extrapolation gives
remarkably accurate total energies (see, Table ~\ref{table:TE}), that are
comparable to or better than those obtained by some popular, sophisticated pure and hybrid density 
functional models\cite{ZB_PRB05}. 
By construction, using the \af\,  values  of this Table,
the MAE in
total energies is same as that in atomization energy. This is 
in contrast to many popular density functional models which seem to 
give better atomization energies due to cancellations of errors 
in total energies of atoms and molecules.
Also, the SR method is unique in that molecules dissociate correctly in the separated 
atom limit. It is therefore not unreasonable to use \af values adjusted for atoms.
\begin{table}
\begin{ruledtabular}
\caption{Mean absolute error (MAE)  in calculated {\it total} energies of G2 
set of 56 molecules for different models. The errors 
are in kcal/mol and are at optimized geometries in respective model.
M1:  SR(EA)/6-311G**/RI-J, M2:  SR(EA)/DGDZVP2/A2, LDA: local density approximation 
(See Ref. ~\onlinecite{ZB_PRB05} for more details)}
 \label{table:TE}
\begin{tabular}{lrdl}
  Model &   MAE \\
 \hline
  M1    &   17.3 \\
  M2    &   16.2  \\
  LDA             &   532 \\
  PBE-GGA         &    101 \\
  B3LYP-hybrid GGA &    15 \\
\end{tabular}
\end{ruledtabular}
\end{table}
The present work shows that fitting perhaps any molecular property can be done 
quantum mechanically through density functional theory.  If the calculations 
are analytic and variational, then small basis sets can be used to generate 
unique, stable and reliable energies using minimal computer time.  Thus 
one can envision embedding quantum-mechanical calculations, with full 
geometry optimization where appropriate, to optimize very sophisticated
quantum-mechanical calculations of molecular properties over the G2 
and larger sets of molecules using PERL scripts to control the 
optimization on a single processor or to farm out independent
sub-optimizations on multiple processors.
Our toolbox of functionals that can be treated analytically contains more\cite{Dunlap00}
than the cube-root functional used in this work.  Thus one can expect that as more experience
is gained in analytic DFT we can better approximate the best exchange and correlation 
functionals that currently require limited-precision numerical integration.  
The process of driving MAEs lower and lower through better and better analytic 
functionals need never end, short of perfect agreement.

        The Office of Naval Research, directly and through the Naval Research Laboratory, and and the 
Department of Defense's   High Performance Computing Modernization Program, through the Common High 
Performance Computing Software Support Initiative Project MBD-5, supported this work.

\end{document}